# Tip and Surface Determination from Experiments and Simulations of Scanning Tunneling Microscopy and Spectroscopy

Óscar Paz,* Iván Brihuega, José M. Gómez-Rodríguez, and José M. Soler†
*Departamento de Física de la Materia Condensada, C-III,*
*Universidad Autónoma de Madrid, E-28049 Madrid, Spain*
(Dated: November 10, 2004)

We present a very efficient and accurate method to simulate scanning tunneling microscopy images and spectra from first-principles density functional calculations. The wave-functions of the tip and sample are calculated separately on the same footing, and propagated far from the surface using the vacuum Green's function. This allows to express the Bardeen matrix elements in terms of convolutions, and to obtain the tunneling current at all tip positions and bias voltages in a single calculation. The efficiency of the method opens the door to real time determination of both tip and surface composition and structure, by comparing experiments to simulated images for a variety of precomputed tips. Comparison with the experimental topography and spectra of the Si(111)-(7×7) surface show a much better agreement with Si than with W tips, implying that the metallic tip is terminated by silicon.

The development of scanning tunneling microscopy [1] (STM) and spectroscopy (STS) has provided an unprecedented knowledge about a rich variety of surface science aspects [2, 3]. STM and STS convey information about the local geometric and electronic structure of metallic and semiconducting surfaces, which has extensively served to unravel their atomic arrangements and reconstructions, but also to analyze surface defects, study adsorbate-covered solid samples or monitor dynamic surface processes like oxidation or diffusion, to give only some examples. The feasibility of atom by atom chemical analysis was dreamed of since an early stage, due to the unique combination of spatial and energetic resolutions. Currently, STM achieves atomic resolution routinely, while discrimination between atomic species has been reported in particular cases. However, a detailed structural analysis, directly from the experiments, is generally far from a minor task, because structural and electronic properties intermixed in the STM images. An even more fundamental difficulty is the lack of control and knowledge on the composition and structure of the tip. This is particularly crucial in the case of STS, where tip states can entirely modify the spectra. As a result of these uncertainties, a careful comparison with theoretical simulations is generally needed to interpret safely the experimental information. To such an end, much progress has been done towards the theory of STM [4] since the pioneering use of perturbation theory by Bardeen [5]. Tersoff and Hamann (TH) [6], made the additional assumption that the tunnel current is dominated by a single *s*-state of the tip, what leads to a simple expression involving only the local density of states (DOS) of the surface. When experiments are highly reproducible, regardless of the tip used, this can be sufficient. However, approaches beyond TH [2, 7, 8], which include the electronic structure of the tip, have been necessary to explain many observations, like bias-dependent images or negative differential resistances. On the other hand, non-perturbative approaches, which treat the interaction between the electrodes more accurately, are needed when the tip is close to contact, but not in the normal tunneling regime of most experiments.

In this Letter we present an efficient perturbative method aimed at real time simulations able to discriminate among different tip and surface structures by comparison with STM and STS experiments. We start from the usual expression [5] for the tunnel current given by the Fermi golden rule,

$$I = \frac{2\pi e}{\hbar} \sum_{t,s} [f(\varepsilon_t) - f(\varepsilon_s)] \, |M_{ts}|^2 \, \delta(\varepsilon_t - \varepsilon_s + eV), \quad (1)$$

where $f(\varepsilon_j)$ is the Fermi–Dirac function, $V$ is the applied bias voltage, and the energies $\varepsilon_t$ and $\varepsilon_s$ are referred to the Fermi levels of tip and sample respectively. $M_{ts}$ is the Bardeen tunneling matrix element between state $\varphi_t$ (with energy $\varepsilon_t$) of the tip and state $\varphi_s$ ($\varepsilon_s$) of the sample,

$$M_{ts} = -\frac{\hbar^2}{2m} \int_\Sigma [\varphi_t^*(\mathbf{r})\nabla\varphi_s(\mathbf{r}) - \varphi_s(\mathbf{r})\nabla\varphi_t^*(\mathbf{r})] \, d^2\hat{\mathbf{r}}, \quad (2)$$

where $\Sigma$ represents any surface located in the vacuum region between both electrodes. The tip and the sample are thus treated as weakly interacting systems with no corrections to the wave-functions and energies of the isolated systems due to their interaction.

The main practical difficulty in evaluating Eq. (2) from density functional theory (DFT) is to obtain sufficiently accurate wave-functions of both tip and sample at the surface $\Sigma$. This difficulty is due to two technical reasons. First, the exponentially decaying wave-functions require an extremely good variational convergence in the vacuum zone, because of their negligible contribution to the total energy. Second, the description of the wave-functions in the vacuum region is limited by the incompleteness of the basis set, specially with bases of atomic orbitals. Since these problems increase with the tip-sample separation, first-principles simulations are frequently performed at unrealistically close separations, thus allowing only for qualitative comparisons. To solve these technical difficulties, we first note that the effective electron potential is nearly flat in the vacuum region, at the typical tunnel distances of 5 − 10 Å, which correspond to currents of $10^{-7} - 10^{-12}$ A (Fig. 1). Then, we use the observation of TH that $M_{ts}$ can be easily obtained if $\varphi_t$ is replaced by the vacuum Green's function $G$, obeying $\nabla^2 G(\mathbf{r}-\mathbf{R}) - \kappa^2 G(\mathbf{r}-\mathbf{R}) = -\delta(\mathbf{r}-\mathbf{R})$, where $\mathbf{R}$ denotes the tip position and $\kappa^2 = \frac{2m}{\hbar^2}(\phi-\varepsilon)$, where $\phi$ is the work function. Substitution into Eq. (2) gives

$$M_{ts}(\mathbf{R}) \propto \int_\Sigma [G^*(\mathbf{r}-\mathbf{R})\nabla\varphi_s(\mathbf{r}) - \varphi_s(\mathbf{r})\nabla G^*(\mathbf{r}-\mathbf{R})]\, d^2\hat{\mathbf{r}} =$$
$$= \varphi_s(\mathbf{R}). \quad (3)$$

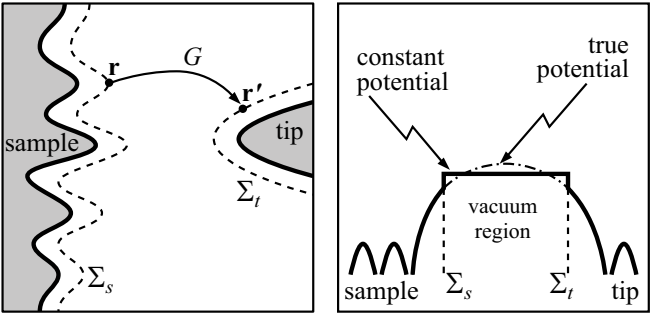

FIG. 1: Schematic view of the propagation of the sample wavefunction through the vacuum region up to the proximities of the tip by means of the vacuum Green's function (left panel). Approximation of a flat effective potential in the vacuum region between the electrodes (right panel).

In the TH approach, Eq. (3) is used to replace $M_{ts}$ by $\varphi_s$ in Eq. (1), eliminating the uncertainties from the tip composition and structure and greatly simplifying the calculation. However, this simplification has its own drawbacks, first because the tip structure does matter in many cases, and second because, as explained previously, $\varphi_s$ is difficult to obtain accurately far from the surface. Therefore, we use Eq. (3) *in the opposite direction*, i.e., to find $\varphi_s$ accurately in the vacuum region. Thus, as illustrated in Fig. 1, the sample states are propagated from a surface $\Sigma_s$ close to the sample, through vacuum, up to another surface $\Sigma_t$ (in the vicinity of the tip), where both wave-functions can be substituted accurately in Eq. (2). Notice that the method is entirely symmetric, and that it can be seen alternatively as the propagation of the tip wave-functions from $\Sigma_t$ to $\Sigma_s$ [9]. A convenient definition of $\Sigma_t$ and $\Sigma_s$ is as isosurfaces of constant electron density, $\rho(\mathbf{r})_{\mathbf{r}\in\Sigma} = \rho_0$. The value $\rho_0$ must be chosen so that the isosurfaces are close enough to the physical surfaces to ensure an accurate description of the first-principles wave-functions, but far enough to make adequate the approximation of a constant effective potential in the vacuum region. In practice, surface integrals can be more efficiently implemented if they are transformed into volume integrals. To this end, we represented $\Sigma$ by the constraint function $S(\mathbf{r}) \equiv \log(\rho(\mathbf{r})/\rho_0) = 0$, so that

$$\int_\Sigma \mathbf{f}(\mathbf{r})\, d^2\hat{\mathbf{r}} = \int \mathbf{f}(\mathbf{r}) \cdot \mathbf{c}(\mathbf{r})\, d^3\mathbf{r}, \text{ with } \mathbf{c}(\mathbf{r}) = \delta(S(\mathbf{r}))\frac{\nabla\rho(\mathbf{r})}{\rho(\mathbf{r})}. \quad (4)$$

After smoothing the delta function (i.e., substituting it by a Gaussian-like function of finite width) the 3D integrals can be performed conveniently in a regular grid. Furthermore, the convolution theorem can be applied in Eq. (3) to express $\varphi_s(\mathbf{r}')$ as an inverse Fourier transform:

$$\varphi_s(\mathbf{r}') = \frac{1}{(2\pi)^{3/2}} \int \tilde{g}(\mathbf{k})\, [\tilde{A}_s(\mathbf{k}) + i\mathbf{k}\cdot\tilde{\mathbf{B}}_s(\mathbf{k})]\, e^{i\mathbf{k}\cdot\mathbf{r}'}\, d^3\mathbf{k}, \quad (5)$$

where $\tilde{g}(\mathbf{k}) = 1/(k^2+\kappa^2)$, $\tilde{A}_s(\mathbf{k})$ and $\tilde{\mathbf{B}}_s(\mathbf{k})$ are the Fourier transforms of $G(\mathbf{r})$, $A_s(\mathbf{r}) = \mathbf{c}_s(\mathbf{r})\cdot\nabla\varphi_s(\mathbf{r})$ and $\mathbf{B}_s(\mathbf{r}) = \mathbf{c}_s(\mathbf{r})\,\varphi_s(\mathbf{r})$. Then, to obtain the tunneling matrix elements, Eq. (5) is substituted into Eq. (2) (shifting tip variables up to $\mathbf{R}$), what leads to

$$M_{ts}(\mathbf{R}) = \frac{\hbar^2}{2m} \int [\tilde{A}_t^*(\mathbf{k}) - i\mathbf{k}\cdot\tilde{\mathbf{B}}_t^*(\mathbf{k})]\, \tilde{g}(\mathbf{k}) \times$$
$$\times [\tilde{A}_s(\mathbf{k}) + i\mathbf{k}\cdot\tilde{\mathbf{B}}_s(\mathbf{k})]\, e^{i\mathbf{k}\cdot\mathbf{R}}\, d^3\mathbf{k}. \quad (6)$$

Finally, in Eq. (1) we broaden the delta function with an empirically-fitted self energy to account for the coupling of the finite calculated systems with their respective bulks. Thus, using fast Fourier transforms for the convolutions, $M_{ts}$ can be evaluated, for all tip positions and bias voltages, in a single computation. The starting point are the values of $\mathbf{c}(\mathbf{r})$, $\varphi_s(\mathbf{r})$, and $\varphi_t(\mathbf{r})$ stored in the points of a uniform grid within the regions of the broadened surfaces $\Sigma_s$ and $\Sigma_t$.

In the following we show simulations for the Si(111)-(7×7) surface and compare them with experimental data. This surface presents a good experimental reproducibility and a rich variety of topographic and spectroscopic characteristics [10], making it an ideal benchmark for STM/STS simulations. The experiments were carried out in ultra-high-vacuum (base pressure below $5 \times 10^{-11}$ Torr) with a home-built variable-temperature STM described elsewhere [11]. Clean reconstructed surfaces were prepared by flashing the samples at 1150°C, after carefully degassing at 600°C for several hours. The samples were then slowly cooled down to room temperature (RT) and transferred to the STM. All the experiments were carried out at RT. For the present measurements we used W tips electrochemically etched from polycrystalline wires. They were cleaned *in-situ*, in ultra-high-vacuum, by heating at temperatures close to 800°C and by field emission against a previously degassed Ta foil (typical voltages and currents were 1 kV and 10 μA). Atomic resolution in STM images, however, was usually obtained after intentional slight tip-sample contacts, which may lead to Si termination of the original W tips. Constant current STM topographies were measured at tunnel currents between 0.1 nA and 2 nA, with sample bias voltages between −2 V and +2 V. Spectroscopic data were taken in the current-imaging-tunneling spectroscopy (CITS) mode [10], which involves acquiring, simultaneously with the constant-current topography, a current vs. voltage ($I - V$) curve, (measured with the feedback loop off) in every point of the surface. Thus, maps of $I(x,y,V)$ are obtained at many different bias voltages $V$, with the tip-sample distance $z(x,y)$ (the topography) determined by a fixed control current and bias voltage.

The CITS maps of $\partial I(x,y,V)/\partial V$ are obtained from $I(x,y,V)$ by direct numerical differentiation [12].

*Ab initio* calculations of the Si(111)-(7×7) surface were previously reported [13, 14]. In the simulations of this work, the surface was mimicked using a repeated slab geometry with four layers of silicon (the lowest of them saturated with hydrogen atoms). Two tips were considered: the first was a tungsten bcc pyramid pointing in the (111) direction, with 20 atoms; the second tip was made of ten silicon atoms [15], in which all dangling-bonds were saturated with hydrogen atoms except that of the apex. The wave-functions of the surface and tips were calculated within DFT in its local density approximation [16] (LDA), using the Siesta code [17, 18]. Core electrons were replaced by norm-conserving pseudopotentials [19], whereas valence electrons were described using a double-$\zeta$ plus polarization (DZP) basis set. A real-space grid with a plane-wave cutoff of 100 Ry was used to perform some integrals and to project the final wave-functions $\varphi_s$ and $\varphi_t$. Only the $\Gamma$ point was used in reciprocal space. The geometries of the surface and tips were relaxed independently until the maximum residual force was below 0.04 eV/Å. The values of the tunnel current $I(x,y,z,V)$, calculated for each tip as explained previously, were dumped on files which were then read by the experimental data-acquisition program [12] and processed in exactly the same way as the experimental data.

Figure 2 shows experimental and theoretical topographic images and profiles. The experimental images show typical features of the empty and occupied states, like the different apparent heights of the faulted and unfaulted half-cells, and the observation of rest-atoms in the occupied states only, which are well reproduced in the simulation with a Si tip. The more quantitative profile comparison with both tips, in the lower panel, shows that the agreement is excellent for the Si tip and considerably worse for the W tip. In the former, the corrugation is appreciably higher, in agreement with the experimental observations that enhanced resolution is usually obtained after intentional tip-surface contacts.

In order to further discriminate between the two proposed tips, Fig. 3 compares the CITS measurements with the simulations using the Si tip. All of the very different experimental patterns, within a window of 2.3 V around the Fermi level, are closely reproduced. This is not the case with a W tip, for which there is a single predominant pattern (not shown) at all negative sample voltages. The difference is specially significant because CITS images at negative voltages sample predominantly the surface electron states at the Fermi level, which feel a lower tunnel barrier, but are very sensitive to the DOS of the tip. Therefore, we can safely conclude that the experimental W tip is in fact terminated with Si, during operation with the highest resolution.

In Fig. 3, it can be observed that the same experimental and simulated images appear at different voltages. In fact, in both spectra there is a range of voltages for which nearly the same pattern is observed, and we have just chosen one representative image of each pattern. This is shown in Fig. 4, where the error bars represent such voltage ranges. It can be seen that the relationship between the experimental and simulated voltages is approximately linear, but with a slope different from one. The explanation for this effect may be related to band-bending effects, or to the underestimation of band gaps in DFT. However, such an explanation is not sufficiently clear at present, and it will require further experiments and simulations.

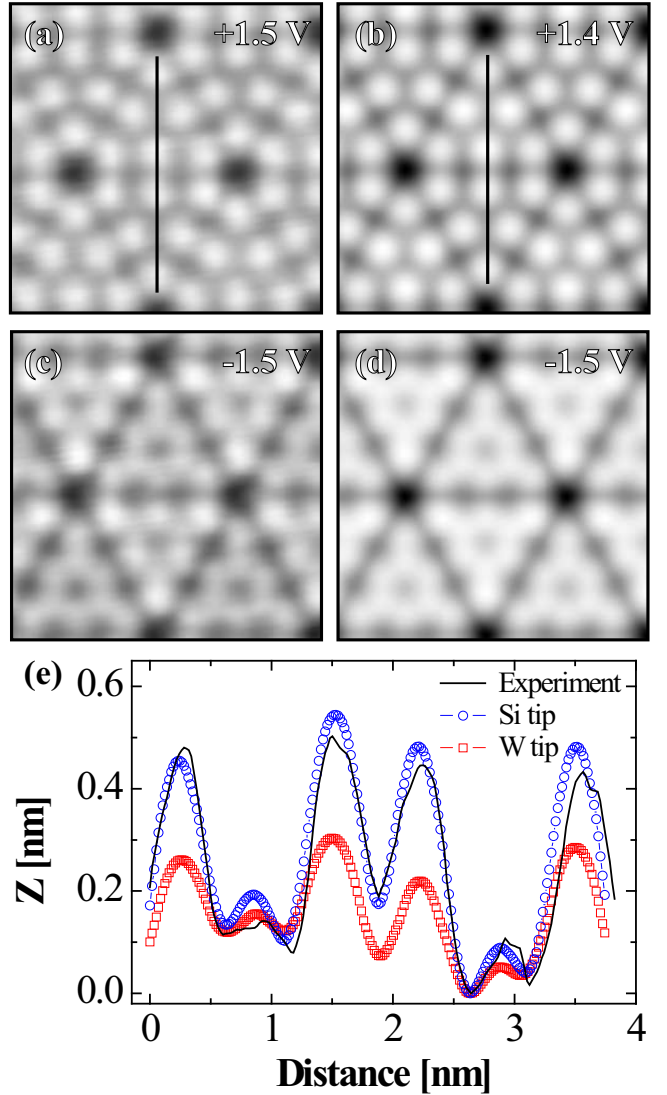

FIG. 2: (Color online) (a), (b) Empty states and (c), (d) occupied states images of the Si(111)-(7×7) surface at 0.2 nA. Left panels show experimental data, while right panels represent simulations with a Si tip, using the same grayscale. (e) Profile comparison between experiment and theory for positive sample voltages, along the solid lines in panels (a) and (b). The three curves have been shifted to make their minima coincide.

In summary, we have presented an accurate method to obtain the tunneling current from first-principles DFT calculations of the surface and the tip at the same footing. The method allows to use efficient basis sets of atomic orbitals, and it provides the current for all tip positions and bias volt-ages in a single calculation, using fast Fourier transforms and convolutions. Despite the very large Si(111)-(7×7) unit cell, the STM calculation (not including the geometry relaxations) takes only ~ 1 h of CPU time in a single processor PC, using a standard DZP basis set. Furthermore, a check with a smaller single-$\zeta$ basis yielded nearly identical images, showing that the quality of the calculated wave-functions is not critical. Thus, the method opens the door to fast comparisons with a variety of precomputed tips, which may allow to interpret STM and STS experiments in real time, determining the structure of both the tip and the surface. In the present case, we have been able to determine that the experimental W tip is in fact terminated with Si.

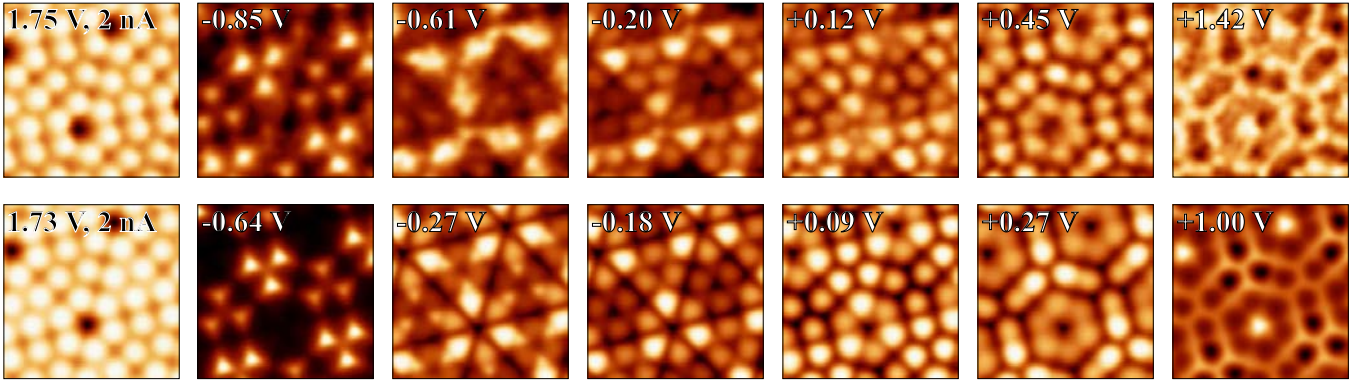

FIG. 3: (Color online) CITS comparison between experiments (top pictures) and simulations using a silicon tip (bottom pictures). The first column shows topographic images at the set-point of 2 nA and ~ 1.75 V. The remaining images represent $\partial I/\partial V$ as a function of $V$ and at the constant tip-sample distance determined by the set-point.

This work has been supported by grants BFM2001-0186, BFM2003-03372 and MAT2001-0664 from Spain's Ministry of Science, and by a doctoral fellowship from the Autonomous Community of Madrid.

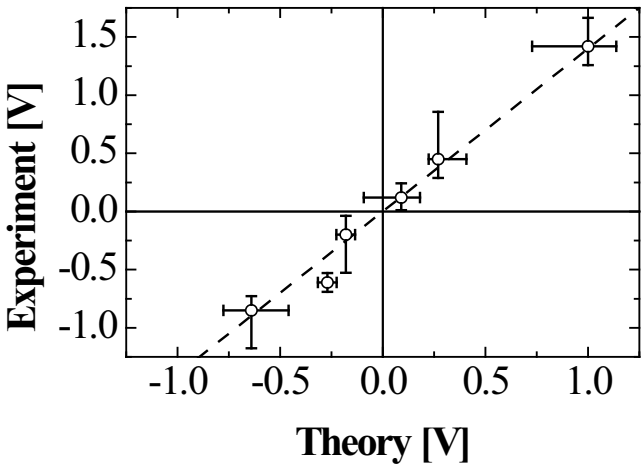

FIG. 4: Bias voltages of the experimental and simulated images shown in Fig. 3. The error bars represent the voltage ranges at which patterns similar to those shown are observed. The dashed line is a linear fit through the origin with a slope of 1.4.